\documentclass[a4paper,11pt]{article}
\usepackage{jinstpub} 
\usepackage{siunitx}

\usepackage{subcaption}

\title{\boldmath Performance of USTC first batch resistive AC-LGAD sensor}

\author[a] {Han Li}
\author[a,1] {Xiao Yang\note{Now at CERN, Esplanade des Particules 1, 1211 Geneva 23, Switzerland}}
\author[a]{Kuo Ma}
\author[a]{Hang Yang}
\author[a]{Aonan Wang}
\author[a]{De Zhang}
\author[a]{Tianao Wang}
\author[a]{Xiangxuan Zheng}
\author[a,2]{Jiajin Ge\note{Now at Department of Physics, University of Michigan, Ann Arbor MI, United States of America
}}
\author[a]{Yusheng Wu}
\author[a]{Hao Liang}
\author[a,3]{Yanwen Liu \note{Corresponding author}}

\affiliation[a] {Department of Modern Physics and State Key Laboratory of Particle Detection and Electronics, University of Science and Technology of China, Hefei 230026, China}

\emailAdd{yanwen@ustc.edu.cn}

\abstract{In this paper, the design and characterization of AC-LGAD sensors at the University of Science and Technology of China is introduced. The sensors are characterized with an infrared laser Transient Current Technique (TCT) system for evaluating signal response characteristics and spatial resolution. The temporal resolution was quantified with electrons emitted by a Sr-90 radioactive source. The spatial resolution can reach 4 $\mu$m and a temporal resolution of 48 ps is achieved.}


\begin{document}
\maketitle
\flushbottom

\section{Introduction}
\label{sec:intro}

The Low Gain Avalanche Diode (LGAD) is a type of n-in-p silicon detector that incorporates a thin, heavily doped gain layer to achieve good timing resolution~\cite{c1}. Through controlled avalanche multiplication within the gain layer, these sensors can achieve a remarkable timing precision of approximately 30 ps. One of the important application of this technology is the upgrade of the ATLAS and CMS experiments at the High-Luminosity Large Hadron Collider (HL-LHC)~\cite{c2}\cite{c3}. However, conventional DC-LGAD design has a fundamental limitation: the presence of no-gain region between adjacent pads due to the segmentation of the gain layer decreases the sensor's fill factor \cite{c4}, which limits the feasibility of finer segmentation.

\begin{figure}[htbp] 
    \centering 
    \includegraphics[width=0.85\textwidth]{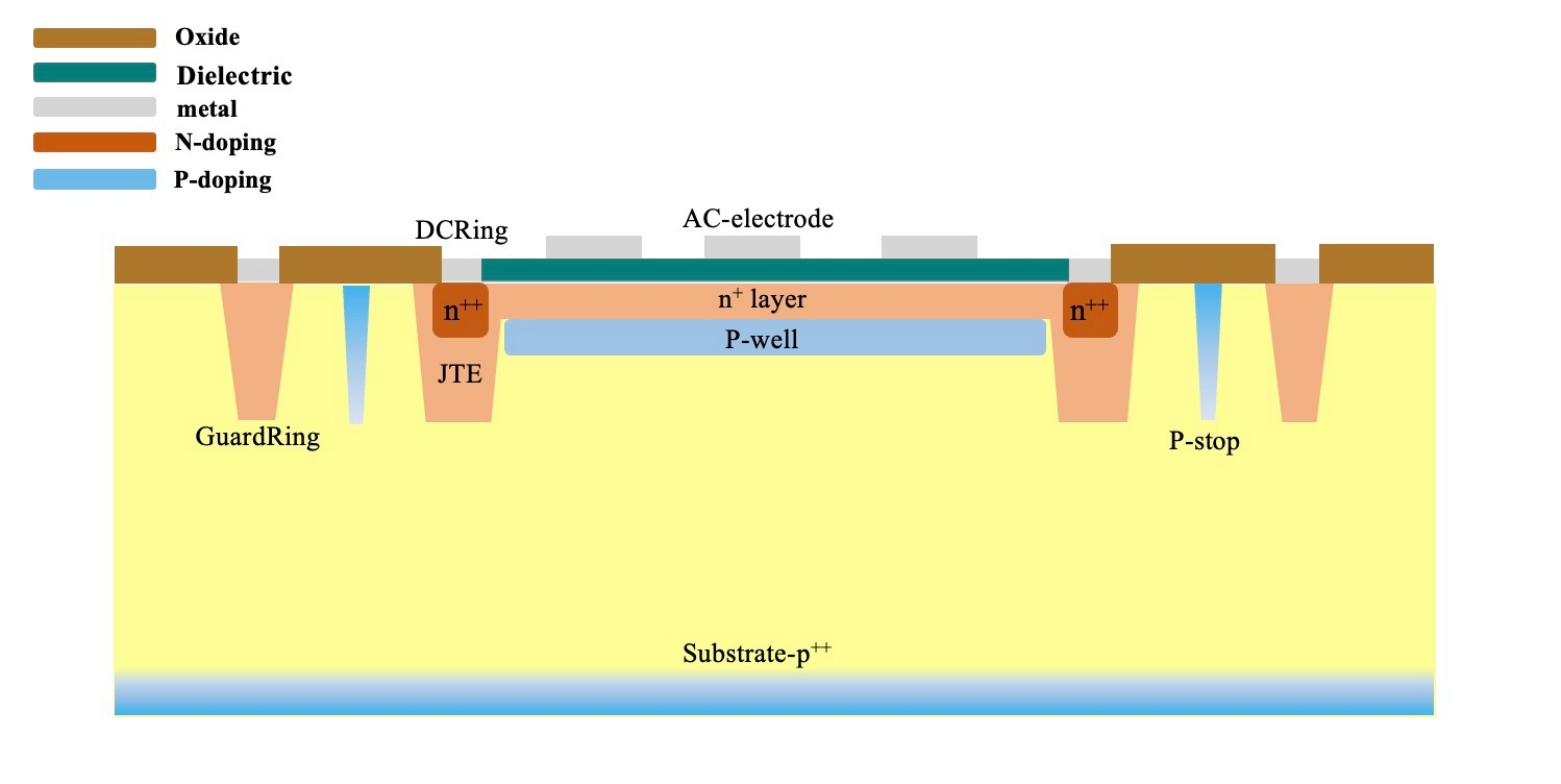}
    \caption{Cross-section of an AC-LGAD device (not-to-scale). From the device physical edge to the center: Guard-Ring, P-stop, JTE, P-well region with AC electrode on top. With a continuous gain layer, the fill-factor issue of DC-LGAD is solved. A dielectric layer between the gain layer and the readout electrode is deposited to achieve the AC-coupled readout.} 
    \label{fig:cross} 
\end{figure}

The resistive AC-LGAD is an innovative development in radiation detector design \cite{c5}\cite{c6}, building upon DC-LGAD while employing an AC-coupled readout method. The cross section of the AC-LGAD is shown in figure \ref{fig:cross}. This configuration features two critical technological advancements. First, a continuous gain implant is used to completely eliminate the no-gain region while maintaining excellent timing resolution inherited from DC-LGAD. Second, a dielectric layer deposited between the gain layer and the readout electrode enables capacitive readout. When a particle traverses the sensor, the electrons and holes from primary ionisation drift towards their respective electrodes. Avalanche multiplication occurs in the gain layer due to the high electric field. The signal is capacitively induced through the dielectric layer to a cluster of readout pads. The charge sharing effect, i.e., an incident particle can cause induced signals on multiple electrodes, can be exploited to improve the spatial resolution. This fast induced signal enables precise temporal reconstruction while maintaining position sensitivity.

This paper presents the comprehensive development of the AC-LGAD detector at the University of Science and Technology of China, including device design, fabrication process optimization, and systematic performance evaluation. The design and fabrication of USTC AC-LGAD will be introduced in section \ref{sec:design}. The sensor characteristics including static characteristics and dynamic properties will be shown in section \ref{sec:signal}. Moreover, the performance including spatial and time resolution of this batch of sensors will be detailed in section \ref{sec:performance}.

\section{Design and fabrication of USTC 1st batch AC-LGAD}
\label{sec:design}

The development is based on the USTC-IME LGAD for the ATLAS HL-LHC upgrade~\cite{c9}. Apart from the standard LGAD fabrication process and common structures, several changes are implemented in the AC-LGAD fabrication process. The $n^{++}$ layer implantation is limited to the area of the DC ring where bias is supplied. The $n^+$ layer is added with a reduced dose to realize a resistive connection. A thin layer of the dielectric material film is deposited for the AC coupling. Fine segmentation is achieved by photolithography to implement electrodes on the metal layer in the active region. 
\begin{figure}[htbp] 
    \centering 
    \includegraphics[width=0.95\textwidth]{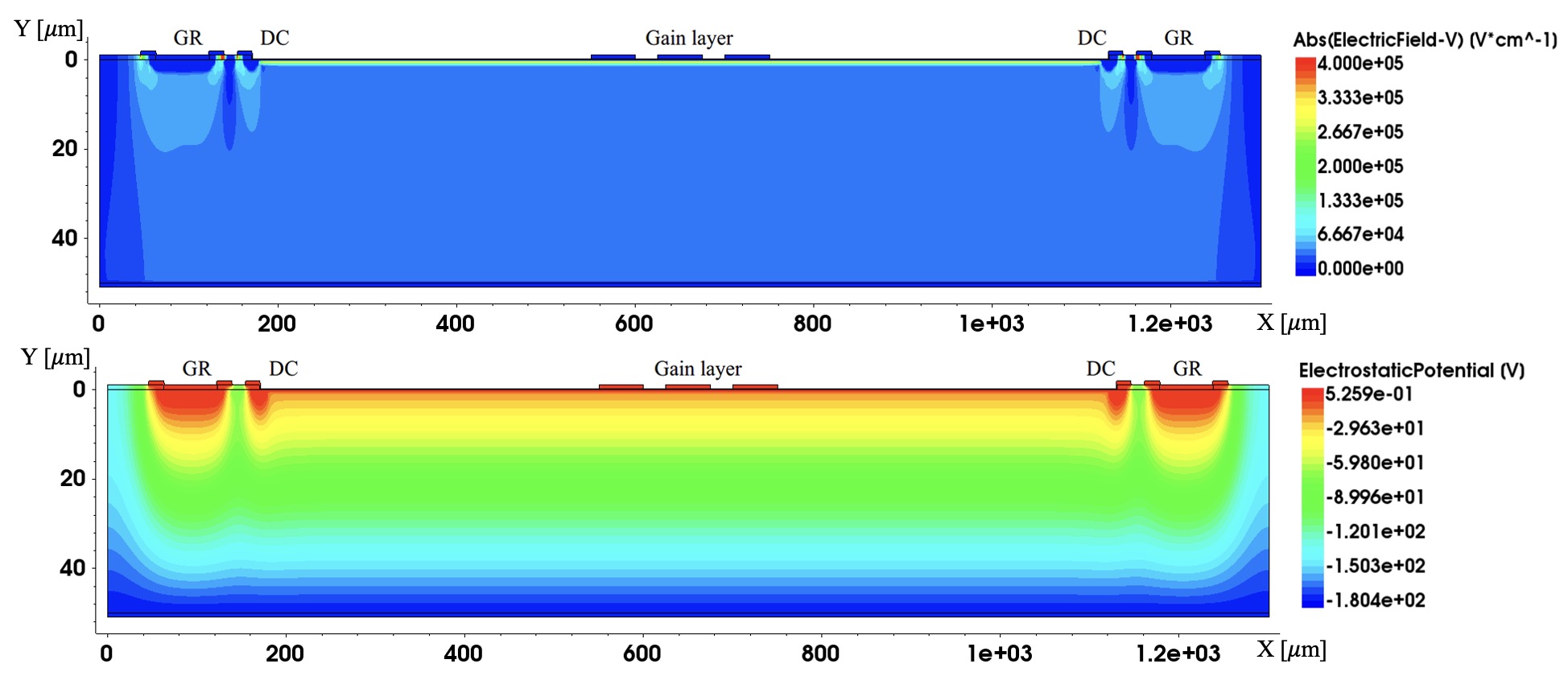}
    \caption{The electric field and potential distribution inside a 50 $\mu$m thick AC-LGAD sensor from the process simulation with a certain process parameter. The structures on the top layer from the edge to the center are: guard ring, DC contact, readout electrodes on top of the gain layer.} 
    \label{fig:TCAD} 
\end{figure}

Based on the previous baseline parameters for the DC-LGAD in Ref~\cite{c9}, the $n^{+}$ implementation, the dielectric layer, and the metal layer pad layout have been re-designed. The optimization is verified with simulation prior to the actual fabrication.

The optimization of AC-LGAD parameters was guided by Technology Computer-Aided Design (TCAD) simulations using $Synopsys Sentaurus^{TM}$ similar to the work in Ref.~\cite{c7} and~\cite{c8}. A multi-physics framework was established to simulate the full fabrication process, including ion implantation, thermal annealing, etc. Figure \ref{fig:TCAD} shows the electric field and potential distribution inside a 50 $\mu$m thick sensor expected for the structure produced with a set of optimized process parameter. The AC-LGAD structure is modeled in the TCAD to study the effects of different process parameters, such as p-stop and GR doping concentrations, on the electrostatic properties of sensors, such as breakdown voltage.

The position-sensitive signal sharing among electrodes is crucial for the AC-LGAD design and depends upon the electrode layout. A simulation tool regarding signal induction and propagation was developed based on the 3D Partial Element Equivalent Circuit (PEEC) solver in the Computer Simulation Technology (CST) Suite \cite{c10}.

\begin{figure}[htbp]
    \centering
    \begin{subfigure}[c]{0.36\textwidth}
        \centering
        \includegraphics[width=\textwidth]{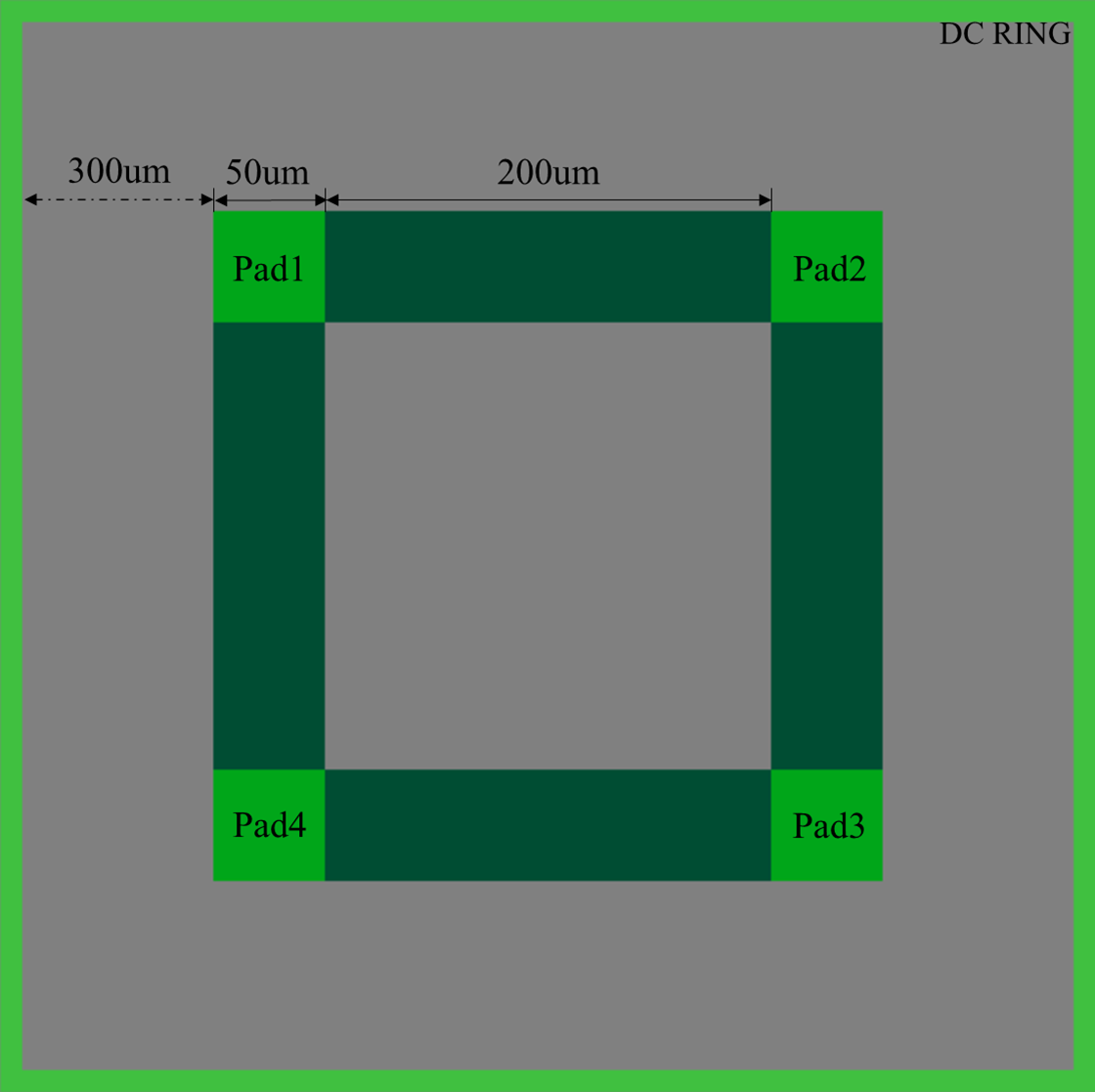} 
        \caption{}
        \label{fig:sim1}
    \end{subfigure}
    \hfill
    \begin{subfigure}[c]{0.56\textwidth}
        \centering
        \includegraphics[width=\textwidth]{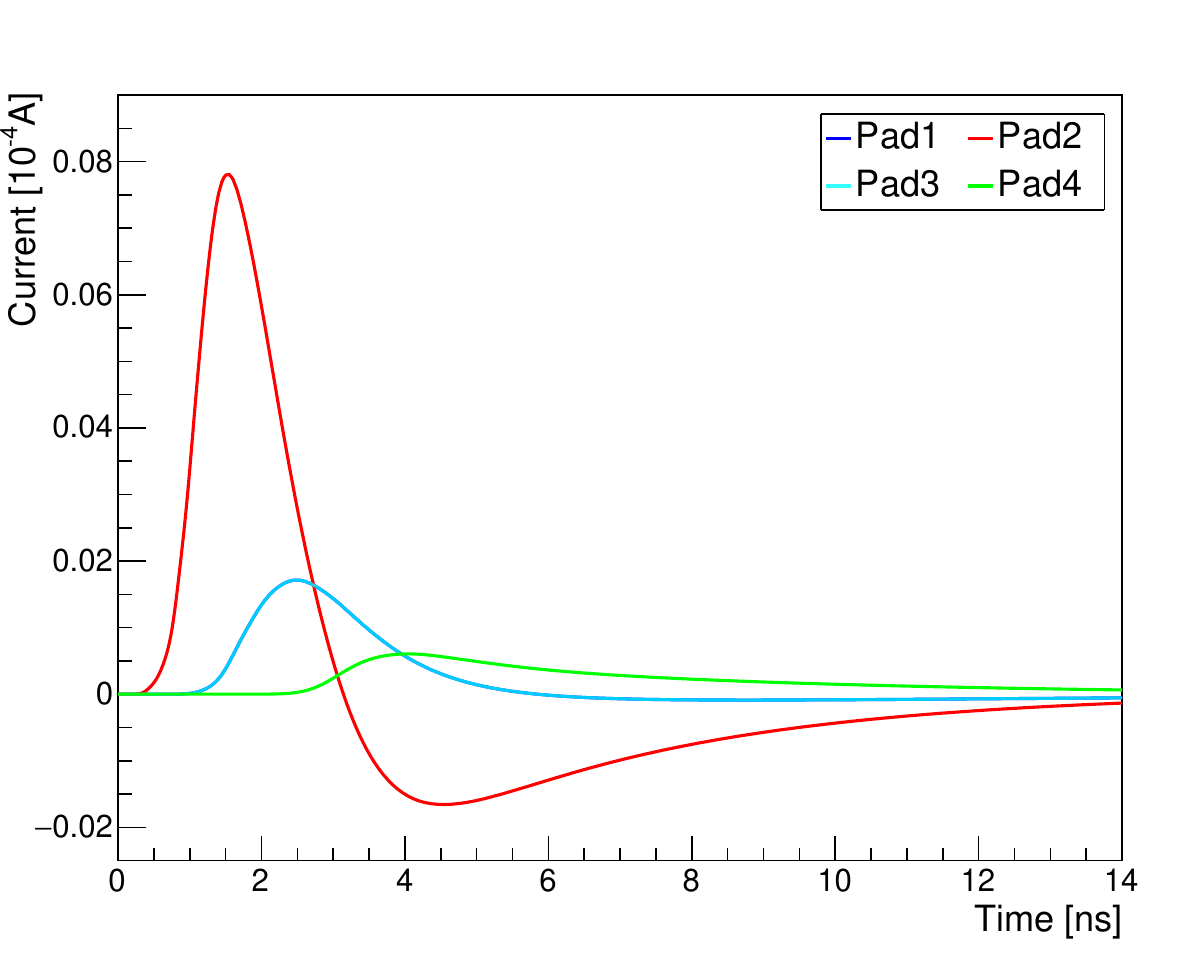} 
        \caption{}
        \label{fig:sim2}
    \end{subfigure}
    \caption{(a) A $2\times2$ pad array setup in CST suite.  The pad size is 50 $\mu$m $\times$ 50 $\mu$m and the pitch size is 300 $\mu$m $\times$ 300 $\mu$m. Virtual pins are created on the readout pads to connect to the amplifiers. Another pin to inject the source current representing the drifting charge carriers. (b) the waveforms observed on the four pads when a source current is injected near the upper right corner as indicated in (a).}
    \label{fig:sim}
\end{figure}

A typical $2\times2$ pad electrode structure is modeled in the PCB studio of the CST Suite, which is shown in figure \ref{fig:sim1}. The PCB studio uses stacked layers to simulate devices, and can simulate different structures of AC-LGAD by setting different conductivity for each layer, such as $n^+$ and dielectric layers. Virtual pins are created on the readout pads to connect to the peripheral electrical elements. The current signals can be injected at designated locations to simulate particle incidence at different positions. Figure \ref{fig:sim2} shows the waveforms observed by a group of four pads when a source current is injected near the upper right corner. Charges are first induced on each electrode, and then discharged to the ground. The rising and falling edges of the signal are related to the resistivity of each layer and the parameters of the peripheral circuit. As expected, the electrode close to the incident position is the first to see the signal and has the highest signal amplitude. 

\begin{figure}[htbp] 
    \centering 
    \includegraphics[width=0.95\textwidth]{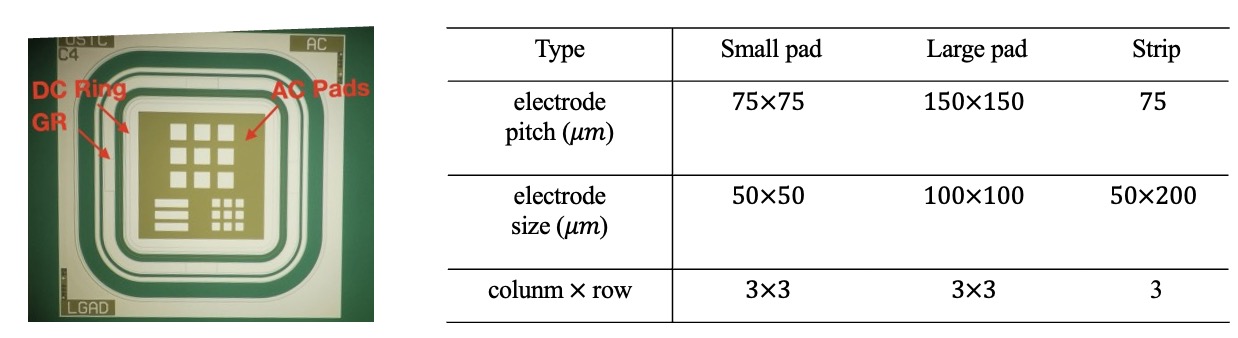}
    \caption{A photo of the USTC AC-LGAD showing the electrode layout of the device (left) and the geometrical parameters of different electrode types: Pad and Strip. The electrode pitch, electrode width, and the numbers of electrodes in column and row directions are summarized (right).} 
    \label{fig:layout} 
\end{figure}

Two wafers with different $n^+$ dose were designed and fabricated at USTC. Wafer number 5 (referred to as W5 hereafter) has a higher $n^+$ dose doping concentration than wafer number 6 (W6). The single sensor size is 1300 $\times$ 1300 $\times$ 50 $\mu$m$^3$, while the active area is 800 $\times$ 800 $\mu$m$^2$. The layout of AC-LGAD sensor used in this study is shown in figure \ref{fig:layout} (left). On each sensor active area, there are three different electrode arrangements. The large pad size is 100 $\times$ 100 $\mu$m$^2$, while pitch is 150 $\mu$m. The small pad size is 50 $\times$ 50 $\mu$m$^2$, while pitch is 75 $\mu$m. The strip electrode size is 200 $\times$ 50 $\mu$m$^2$, while pitch is 75 $\mu$m. The electrode pitch, electrode width, and the numbers of electrodes in column and row directions are summarized in figure \ref{fig:layout} (right).

\section{Sensor characteristics}
\label{sec:signal}

\subsection{Static characteristics}
The static characteristics of the sensors were tested to examine the impact of different process parameters on the devices. To this aim, Current-Voltage (IV) and Capacitance-Voltage (CV) measurements on probe station have been performed at room temperature, as these curves can give useful information, such as breakdown voltage ($V_{\textrm{BD}}$) and depletion voltage of gain layer ($V_{\textrm{GL}}$).

\begin{figure}[htbp]
    \centering
    \begin{subfigure}[b]{0.49\textwidth}
        \centering
        \includegraphics[width=\textwidth]{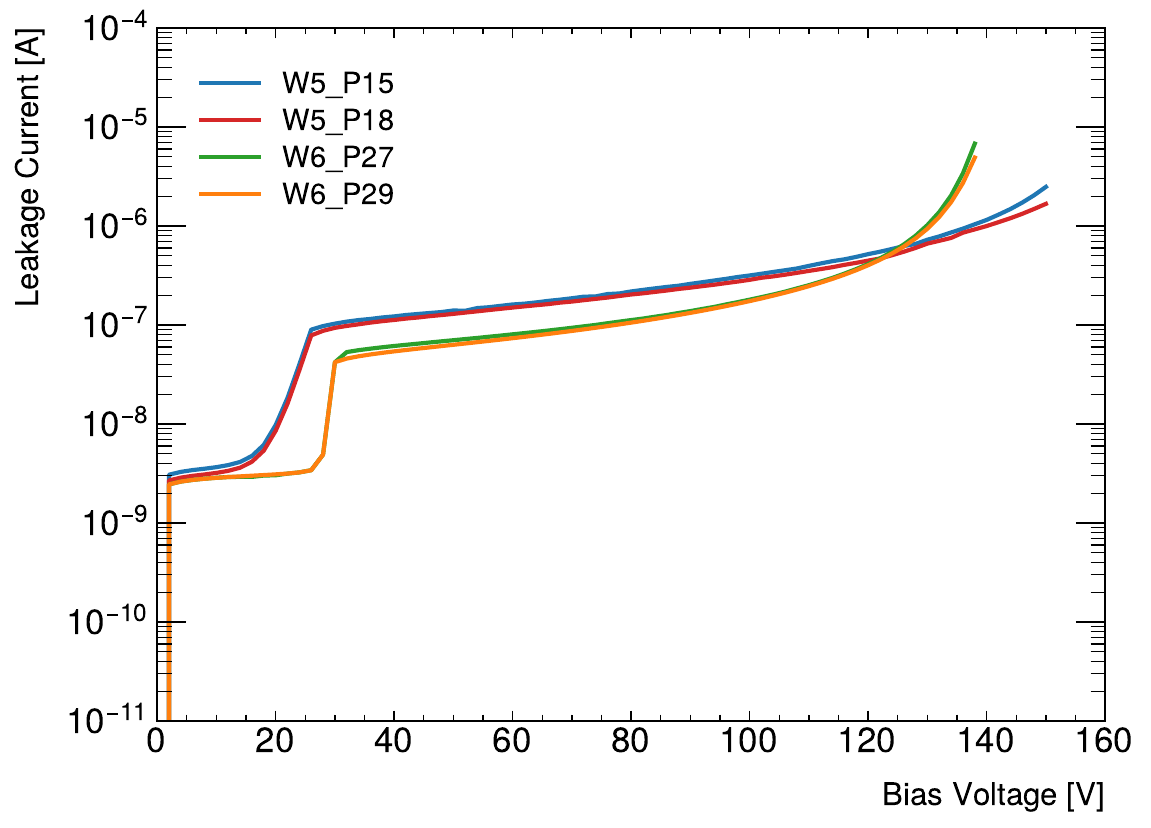} 
        \caption{}
        \label{fig:IV}
    \end{subfigure}
    \hfill
    \begin{subfigure}[b]{0.49\textwidth}
        \centering
        \includegraphics[width=\textwidth]{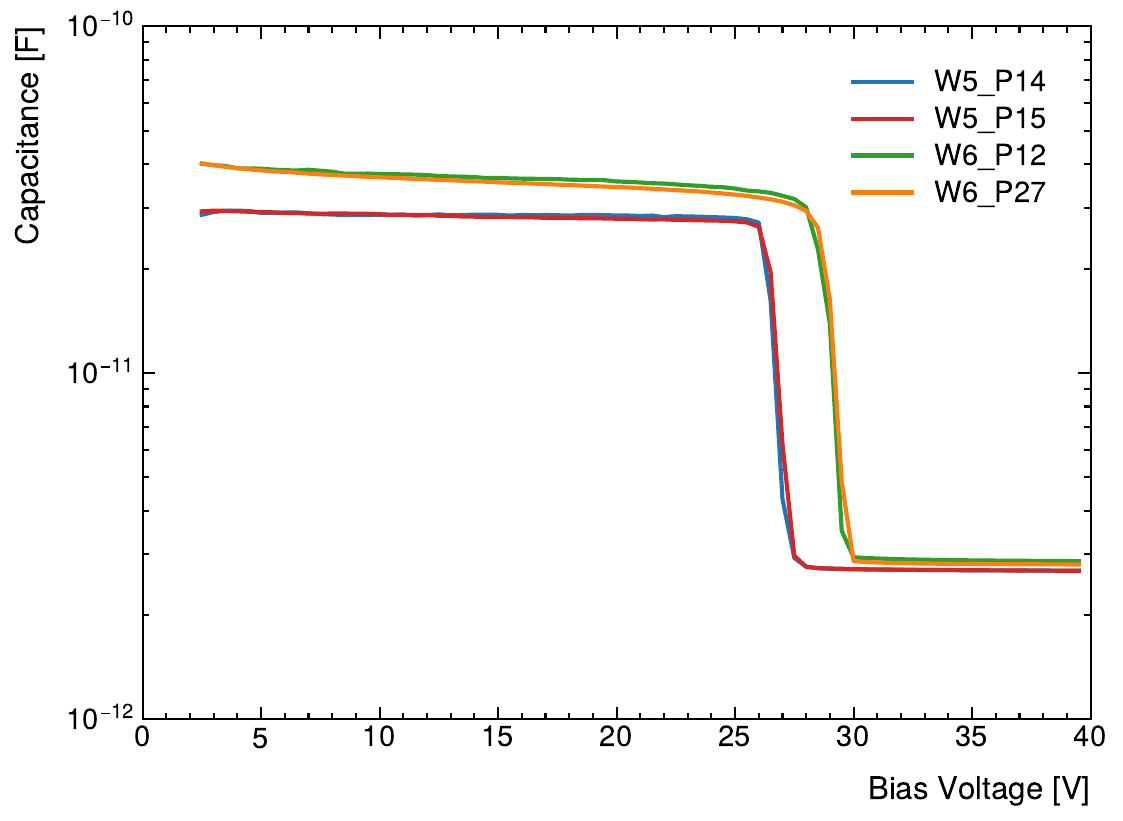} 
        \caption{}
        \label{fig:CV}
    \end{subfigure}
    \caption{IV and CV curves of the AC-LGAD sensors fabricated at USTC. These samples come from different positions on two wafers while W represents the wafer label and P represents the position label.}
    \label{fig:probe}
\end{figure}

Figure \ref{fig:probe} presents I(V) and C(V) characteristics of several devices fabricated on W5 and W6. The observed breakdown voltage is defined as the voltage corresponding to a current of 10 $\mu$A. the $V_{\textrm{BD}}$ of W5 is about 145 V, while W6 is relatively lower at 135 V. The difference in IV curves and the $V_\textrm{BD}$ between sensors from W5 and W6 is caused by the difference in $n^+$ layer implantation concentration. The $V_{\textrm{GL}}$ of W5 extracted from CV curve is about 27 V, while the $V_{\textrm{GL}}$ of W6 is relatively highly at 30 V. The $V_\textrm{GL}$ extracted by IV and CV curves is consistent, and the depletion capacitance from samples W5 and W6 is around 3 pF.

\subsection{Dynamical properties}

The sensor's signal properties were evaluated with an infrared (IR) laser transient current technique (TCT) system through localized charge injection.

An IR laser TCT system was set up to stimulate the production of charge carriers and readout induced signals from detectors. The wavelength of IR laser is 1064 nm, and after focusing with the lens group, the diameter of beam spot is less than 10 $\mu$m. A stage platform with a spatial precision better than 1 $\mu$m used for the movement in X-Y-Z dimensions. Four pads are wire bonded to the USTC 9-channel amplifier for readout \cite{c11}. The signal pulses from four AC pads are recorded by a four channels oscilloscope with 1 GHz bandwidth and 20 GS/s sampling rate for offline analysis.

\begin{figure}[htbp]
    \centering
    \begin{subfigure}[b]{0.49\textwidth}
        \centering
        \includegraphics[width=\textwidth]{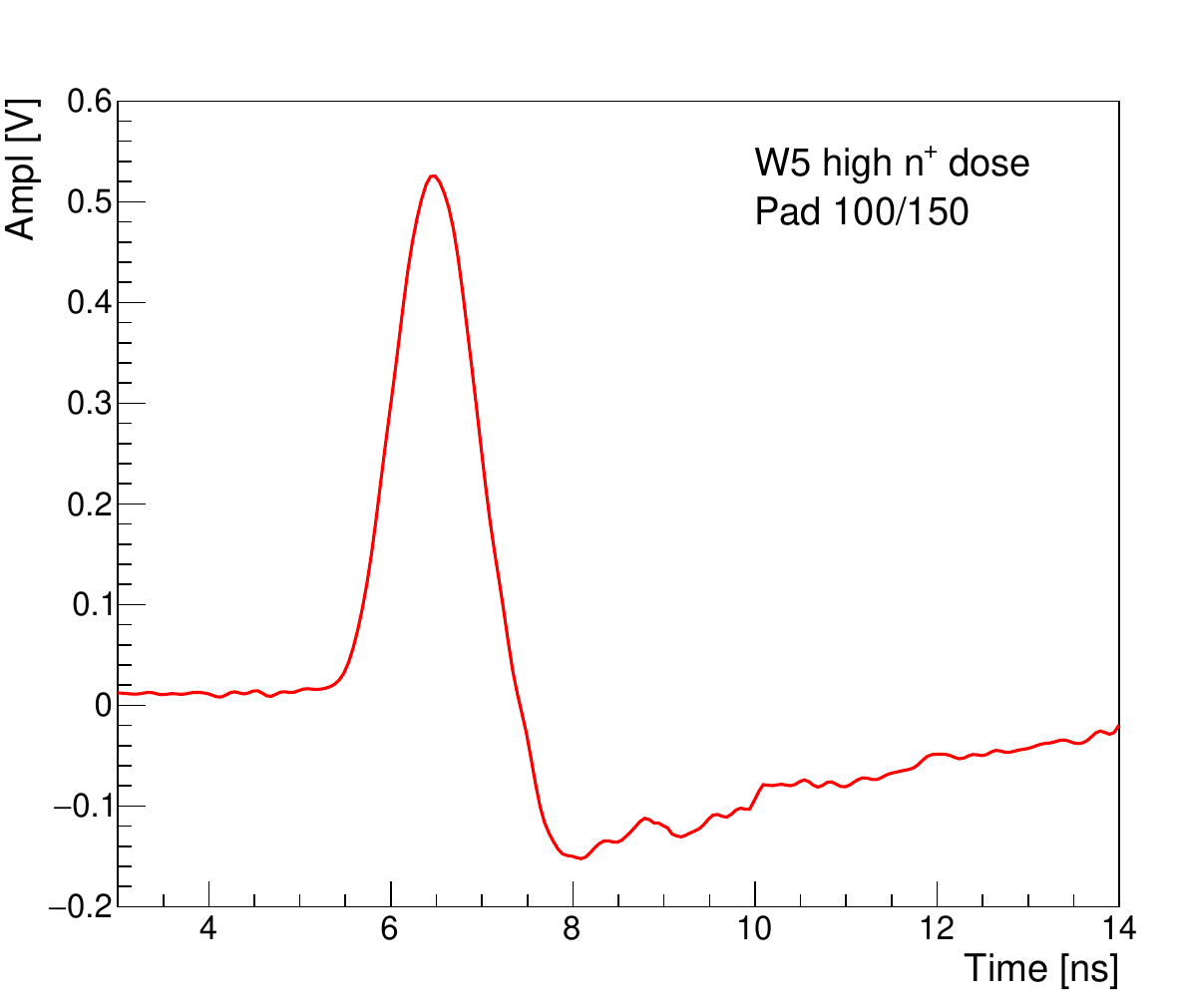} 
        \caption{}
        \label{fig:wave1}
    \end{subfigure}
    \hfill
    \begin{subfigure}[b]{0.49\textwidth}
        \centering
        \includegraphics[width=\textwidth]{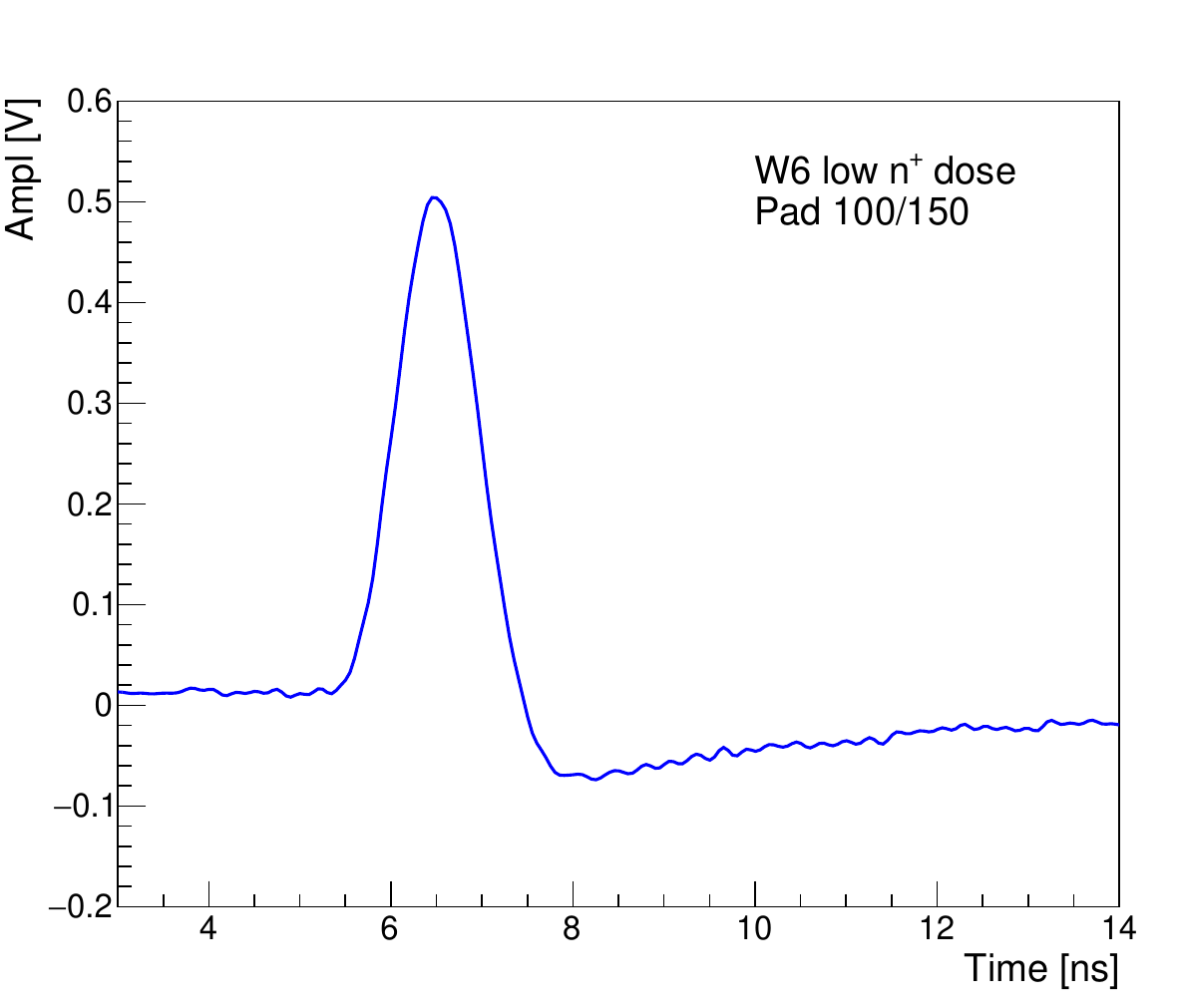} 
        \caption{}
        \label{fig:wave2}
    \end{subfigure}
    \caption{Waveform acquired with the TCT setup from 100/150 pad electrode on W5 (a) and W6 (b), the nature of bipolar can be clearly seen. Since W5 has a higher $n^+$ layer doping concentration, the undershoot is more visible in its waveform.}
    \label{fig:wave}
\end{figure}

The laser signals from the pixel electrode region of the sensors on two wafers acquired with TCT system are shown in figure \ref{fig:wave}. Since the overall integrated charge in AC pads is zero, the nature of bipolar can be clearly seen. Another interesting aspect is that different RC constant produce different signal shape. The waveform from W5 exhibits more overshoot, as its higher $n^+$ layer doping concentration results in a smaller RC constant.

\begin{figure}[htbp]
    \centering
    \begin{subfigure}[b]{0.49\textwidth}
        \centering
        \includegraphics[width=\textwidth]{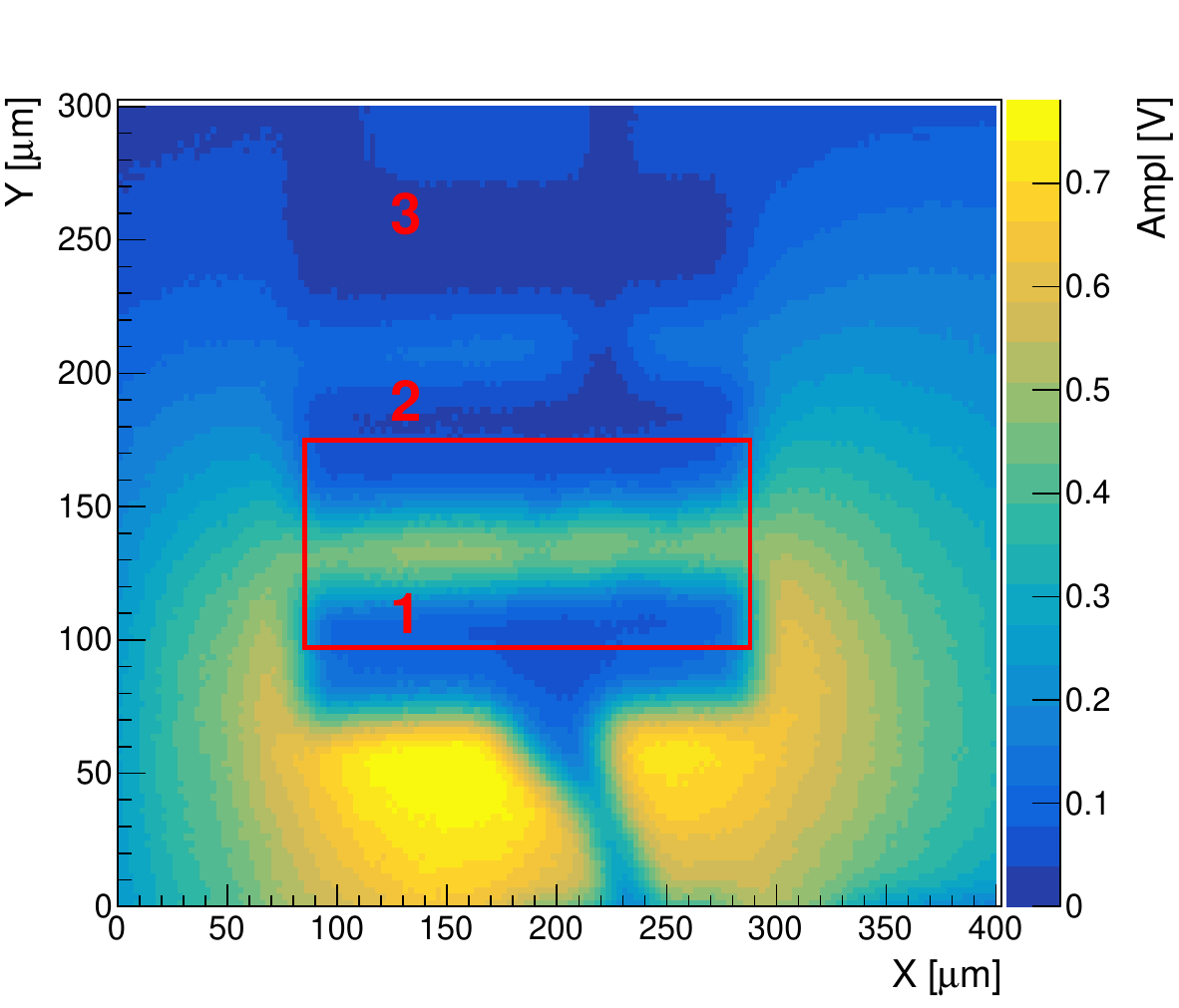} 
        \caption{}
        \label{fig:2D_strip}
    \end{subfigure}
    \hfill
    \begin{subfigure}[b]{0.49\textwidth}
        \centering
        \includegraphics[width=\textwidth]{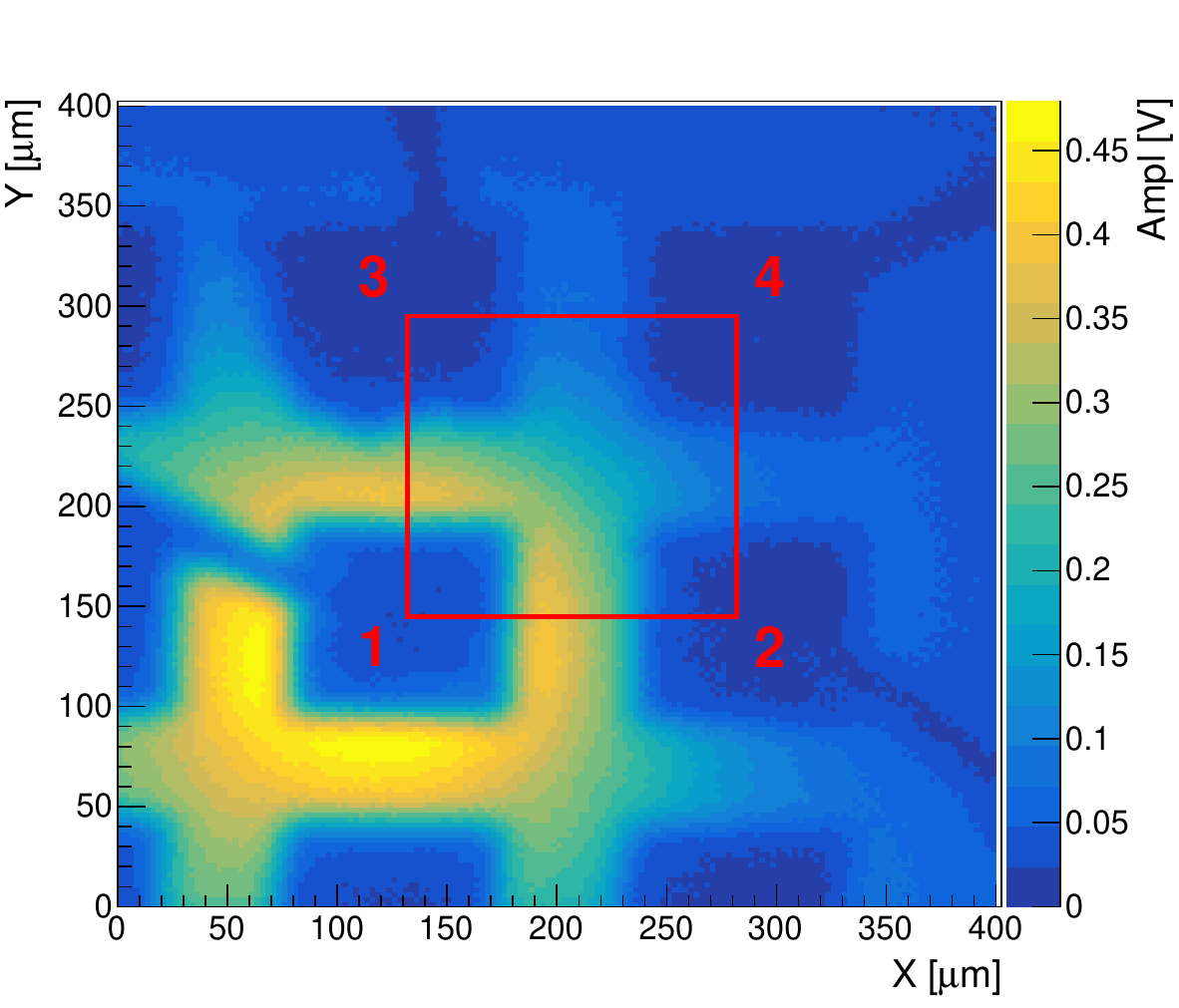} 
        \caption{}
        \label{fig:2D_pad}
    \end{subfigure}
    \caption{The response map to the laser on the sensor active area surface. The XY represents laser incidence position, while the Z-axis represents signal peak amplitude. The indices of the electrodes are indicated in the plot. The red rectangle denotes the region of interest for subsequent analysis of performances. (a): the signal induced on the strip 1 in the strip array (b): the signal induced on the pad 1 in the large pad array.}   \label{fig:2D}
\end{figure}

\begin{figure}[htbp]
    \centering
    \begin{subfigure}[b]{0.49\textwidth}
        \centering
        \includegraphics[width=\textwidth]{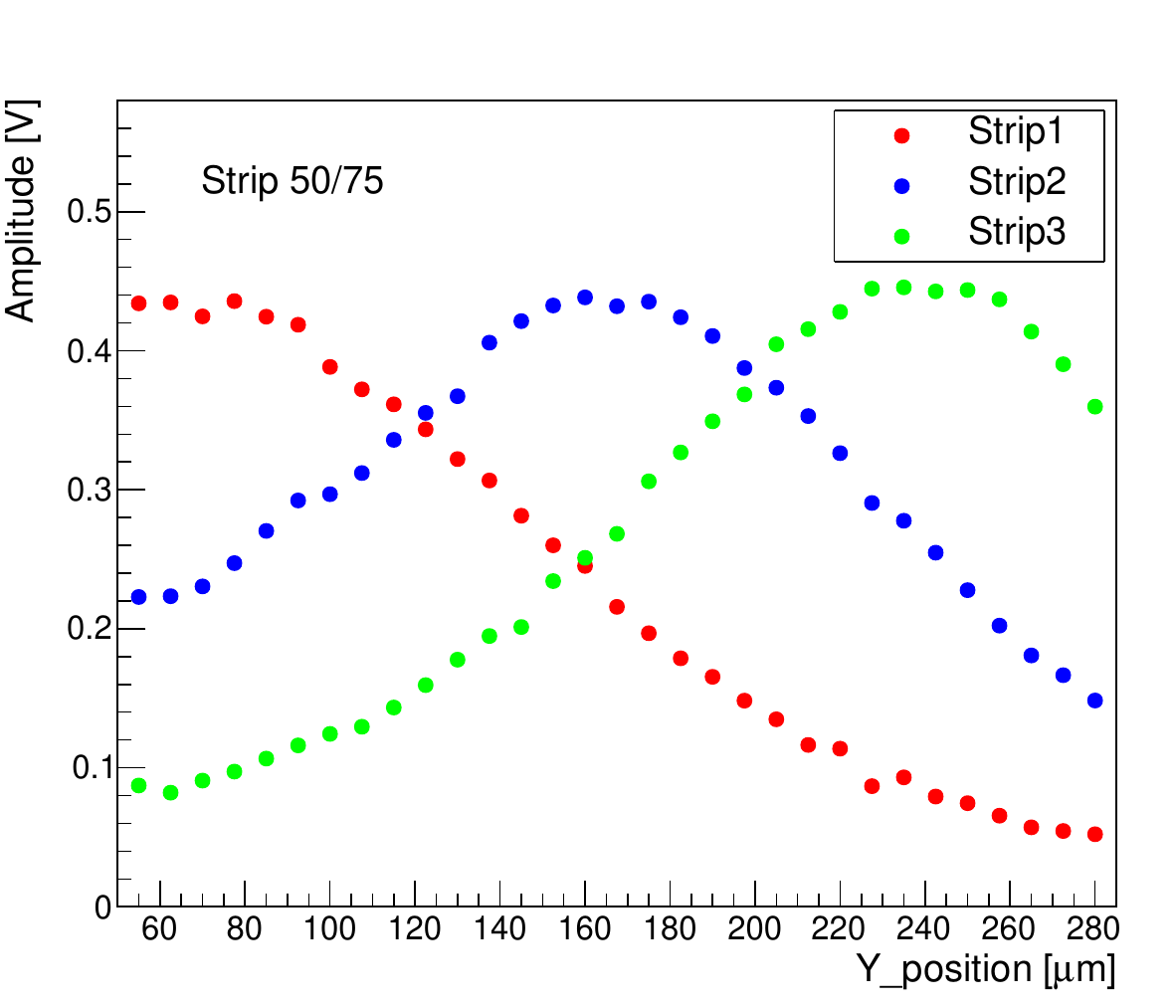} 
        \caption{}
        \label{fig:strip}
    \end{subfigure}
    \hfill
    \begin{subfigure}[b]{0.49\textwidth}
        \centering
        \includegraphics[width=\textwidth]{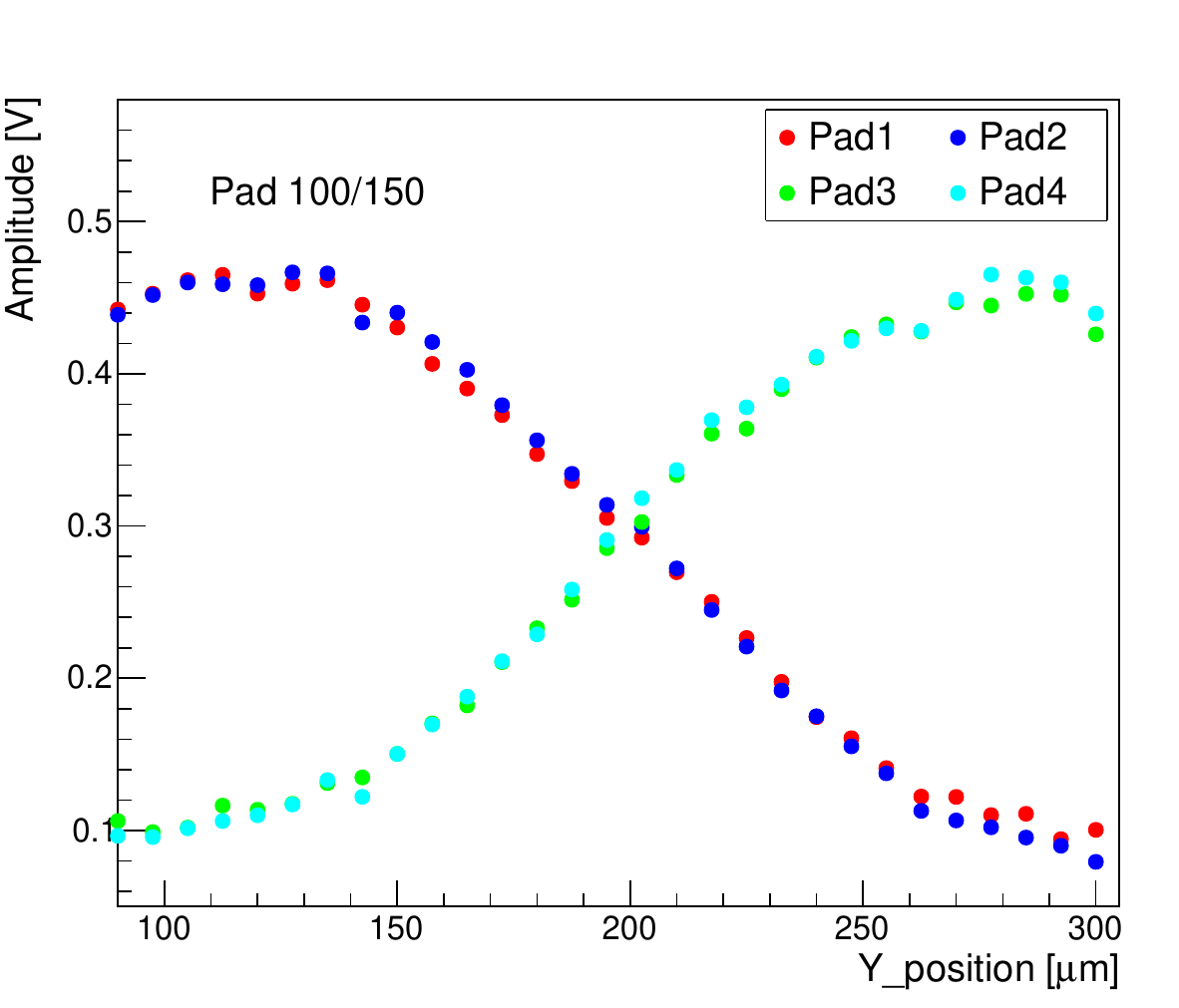} 
        \caption{}
        \label{fig:pad}
    \end{subfigure}
    \caption{The signal amplitude as a function of the hit position along the y-direction. (a): the variation of signal amplitude on the three strip electrodes at X = 50 $\mu$m. The electrode numbering sequence is from bottom to top (see figure \ref{fig:2D_strip}). (b): the variation of signal amplitude on the four pad electrodes at X = 200 $\mu$m. The electrode numbering sequence is from bottom left to top right (see figure \ref{fig:2D_pad}).}
    \label{fig:curve}
\end{figure}

\begin{figure}[htbp]
    \centering
    \begin{subfigure}[b]{0.49\textwidth}
        \centering
        \includegraphics[width=\textwidth]{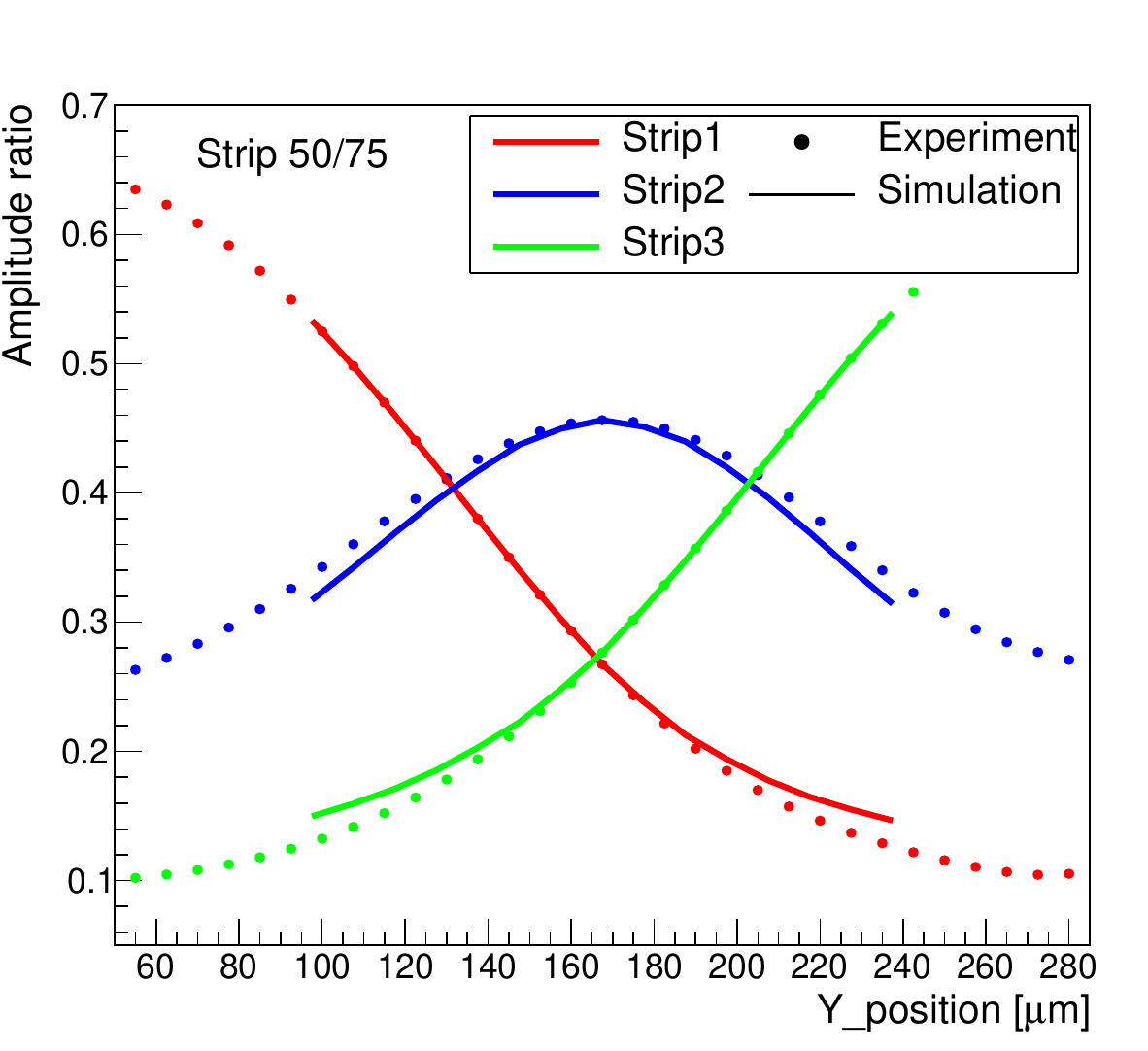} 
        \caption{}
        \label{fig:sim_strip}
    \end{subfigure}
    \hfill
    \begin{subfigure}[b]{0.49\textwidth}
        \centering
        \includegraphics[width=\textwidth]{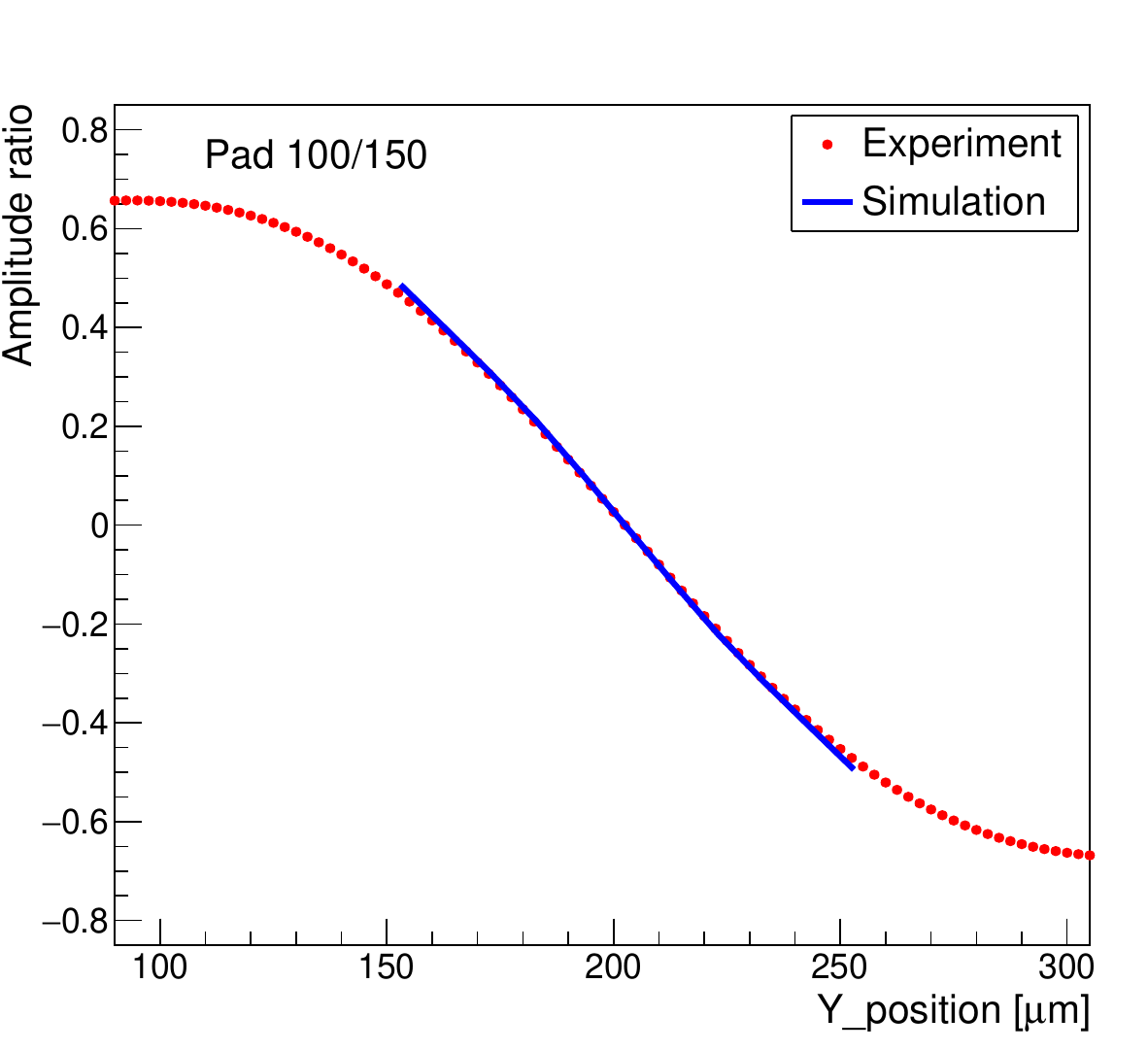} 
        \caption{}
        \label{fig:sim_pad}
    \end{subfigure}
    \caption{The comparison between laser test and CST simulation result.  The dots represents the experimental results, the line represent the simulation results.  The simulation is done by a step size is 10 $\mu$m and is performed in the range of the pitch size. There is good consistency between simulation and experiment within the pitch range.}
    \label{fig:sim_exp}
\end{figure}

By utilizing the three-dimensional stage platform, the response of the sensors to laser incidence at different positions can be obtained. The signal amplitudes on a given electrode as a function of hit positions are shown in figure \ref{fig:2D} for AC-LGAD from W5 with strip and pad electrodes. For the laser scanning of the strip type AC-LGAD, the 3 adjacent strips are wired-bonded to the amplifier board. For the pad type AC-LGAD, the 4 pads in a 2 $\times$ 2 array are wire-bonded to USTC multi-channel board. The metal electrodes and bonding wires blocked the IR laser, causing the ``shades'' in the response map. The sensor surface was scanned by steps of 2.5 $\mu$m. The signal amplitude varies with the distances to the electrodes due to charge sharing. The signal amplitude on each electrodes for a series of hit position along the y-direction are shown in figure \ref{fig:curve}. This kind of charge sharing effects can be exploited to reconstruct the hit positions to improve the spatial resolution. 

The proportion of signal amplitude on each electrode was compared between laser test and simulation result from CST, as shown in figure \ref{fig:sim_exp}. The amplitude ratio $R_{\textrm{strip}\:i}  = \frac{A_i}{\sum_{j=1}^{n} A_j}$ and $R_\textrm{pad} = \frac{A_1+A_2-A_3-A_4}{A_1+A_2+A_3+A_4}$ are used for verification, respectively. The comparison shows good consistency within the pitch range, verifying the reliability of the simulation software.

\section{Spatial and temporal resolution}
\label{sec:performance}

The pulsed laser is used to generate AC signals in several positions on the detector surface. The waveform produced by laser injection was recorded by oscilloscope for position reconstruction at room temperature. A $^{90}$Sr source is used to test the sensor's timing performance at 20$^\circ$C. Both spatial and temporal resolutions are evaluated. 

\subsection{Spatial resolution}

Before the reconstruction process, two preparatory steps are carried out. Firstly, the responses of the 4 pad electrodes are equalized to get rid of imbalances caused by the setup (different lengths of wires, different gains in amplifiers, etc) that can lead to biases. To achieve this, a group of 4 electrodes is selected and laser is injected to the geometric center of the cluster. The amplitudes of the 4 electrodes are scaled so that the average of them are the same. Secondly, the laser intensity is tuned so that the amount of introduced charges is consistent with that of Minimum Ionising Particle.

The hit position is reconstructed with the amplitude ratio method. Two ratios $F_x$ and $F_y$ are defined in two directions respectively as in eq(~\ref{eqFraction}). The $F_x$ and $F_y$ are strong indicators of $x$ and $y$ coordinates of the hit position. With a training dataset from the IR laser test, lookup tables of $F_x : x$ and $Fy:y$ are constructed and is later used to reconstruct the hit positions.  
        \begin{equation}
        F_x = \frac{A_1+A_3-A_2-A_4}{A_1+A_2+A_3+A_4}       \quad\quad
        F_y = \frac{A_1+A_2-A_3-A_4}{A_1+A_2+A_3+A_4}
        \label{eqFraction}
        \end{equation}

\begin{figure}[htbp]
    \centering
    \begin{subfigure}{0.48\textwidth}
        \includegraphics[width=\linewidth]{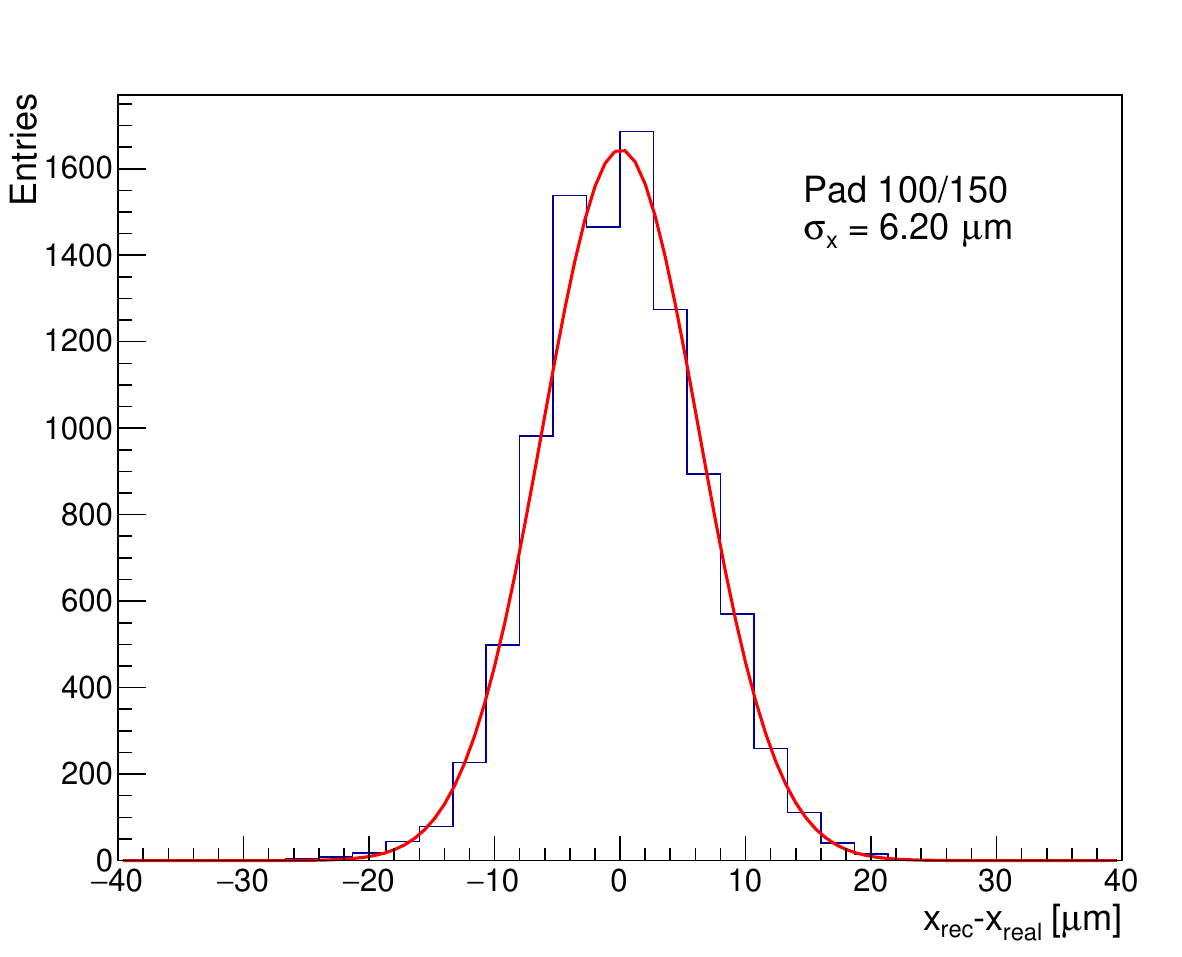}
        \caption{}
        \label{fig:sub1}
    \end{subfigure}
    \hfill
    \begin{subfigure}{0.48\textwidth}
        \includegraphics[width=\linewidth]{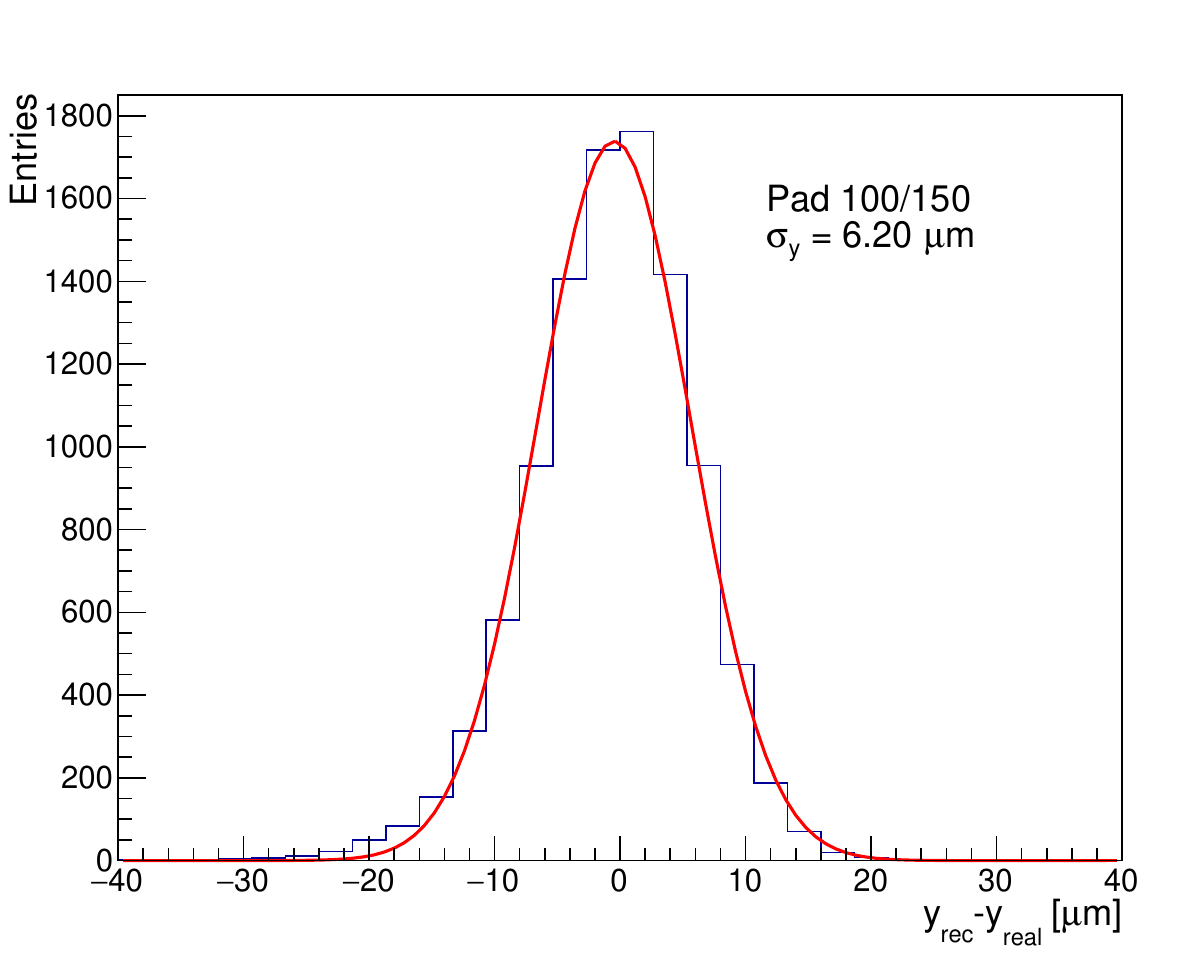}
        \caption{}
        \label{fig:sub2}
    \end{subfigure}

    \begin{subfigure}{0.48\textwidth}
        \includegraphics[width=\linewidth]{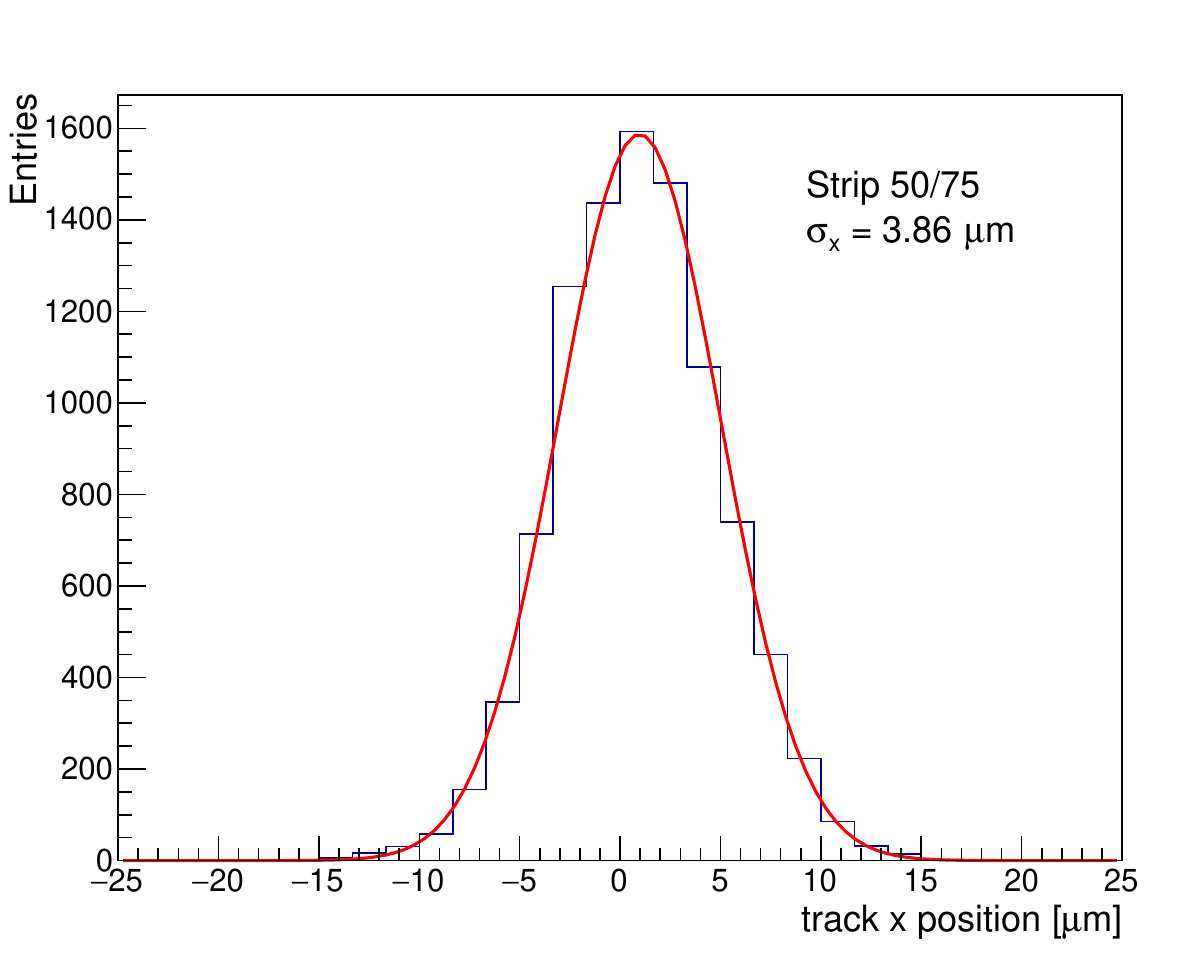}
        \caption{}
        \label{fig:sub3}
    \end{subfigure}
    \hfill
    \begin{subfigure}{0.48\textwidth}
        \includegraphics[width=\linewidth]{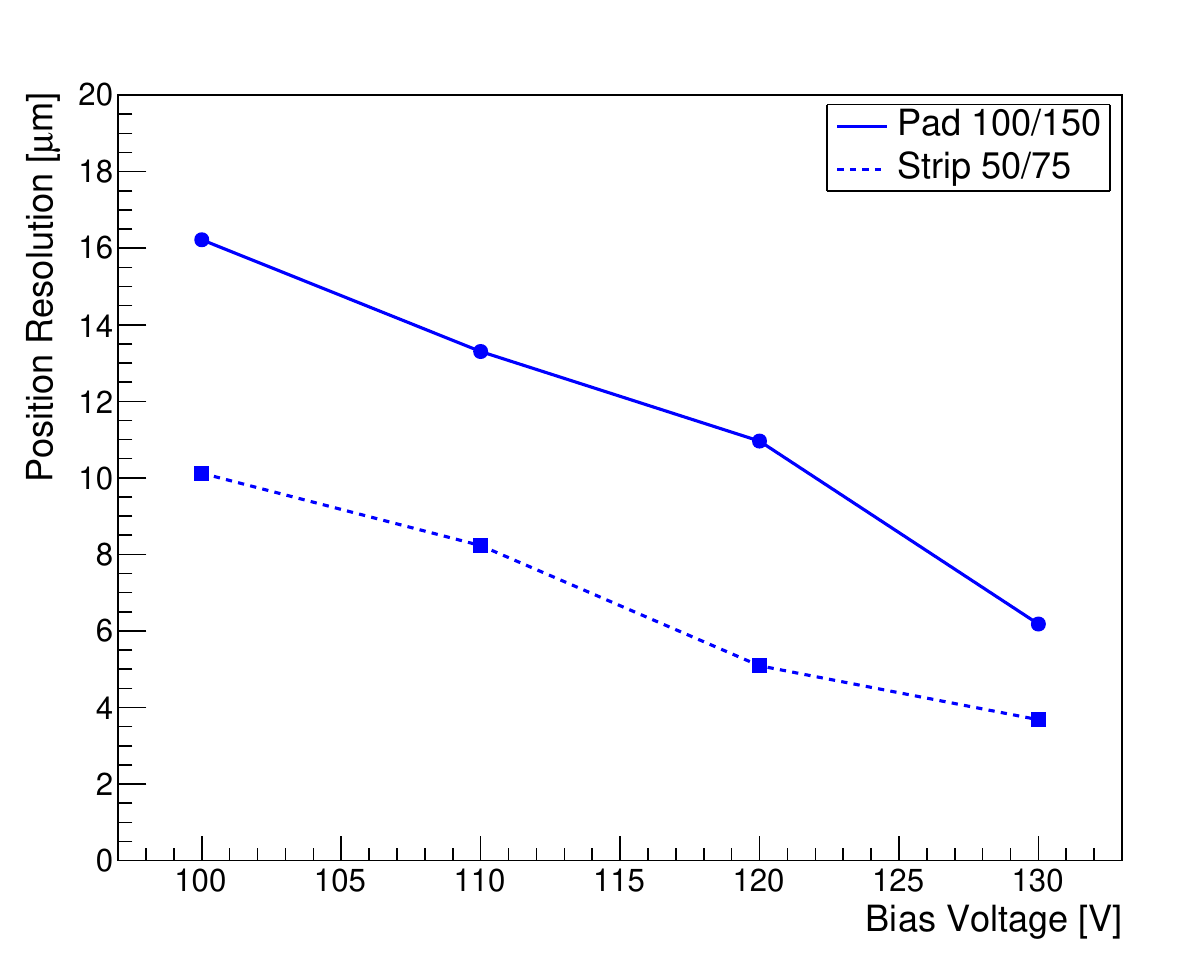}
        \caption{}
        \label{fig:sub4}
    \end{subfigure}
    \caption{(a): Distribution of the residual of the reconstructed positions for 100/150 (size/pitch in $\mu$m) pad electrode overlaid with Gaussian fits in the x direction. (b): distribution of the residual the reconstructed positions for pad in the y direction. (c): residual of the reconstructed positions for 50/75 strip. (d): relationship between position resolution and applied bias voltage.}
    \label{fig:12}
\end{figure}

Figure \ref{fig:12} shows the distributions of the reconstructed ratios (respectively $x_\textrm{rec}$ and $y_\textrm{rec}$), obtained from a 4-pad cluster configuration on 150-$\mu$m-pitch samples frow W5, overlaid with Gaussian fits. For event selection, the region of interest is shown in Figure \ref{fig:2D}. Spatial resolution of 6.2 $\mu$m is demonstrated. For the position reconstruction of strip electrodes, the workflow is similar. The charge sharing information is presented by the amplitude ratio of $F =  \frac{A_1}{A_1+A_2}$. The $A_1$ and $A_2$ represent the peak amplitudes of the signals observed on two strips. The spatial resolution of 50/75 (size/pitch in $\mu$m) strip electrodes are shown in figure \ref{fig:sub3}. In figure \ref{fig:sub4}, the spatial resolution for strip and pad electrode is shown as a function of the bias voltage. The spatial resolution improves from about 10 $\mu$m to 4 $\mu$m for strip as the bias voltage is increased. In terms of position resolution, the W6 devices have almost identical performance to the W5 device.

For the laser intensity equivalent to MIP, both the strip and pad layouts provide position resolution significantly superior to pitch/$\sqrt{12}$, demonstrating the advantage of charge sharing.
\subsection{Temporal resolution}

\begin{figure}[htbp]
    \centering
    \begin{subfigure}[c]{0.48\textwidth}
        \centering
        \includegraphics[width=\textwidth]{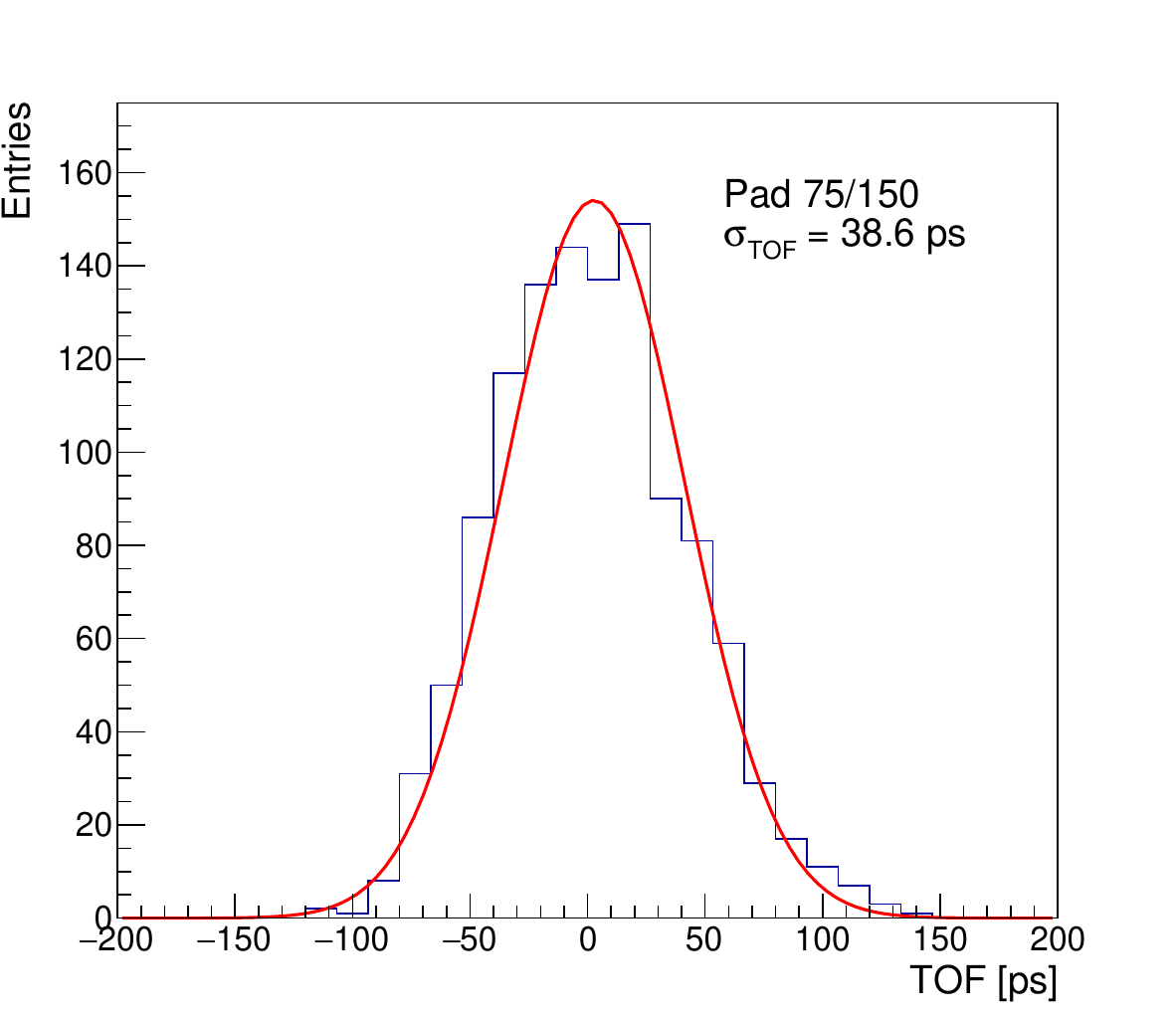} 
        \caption{}
        \label{fig:1channel}
    \end{subfigure}
    \hfill
    \begin{subfigure}[c]{0.48\textwidth}
        \centering
        \includegraphics[width=\textwidth]{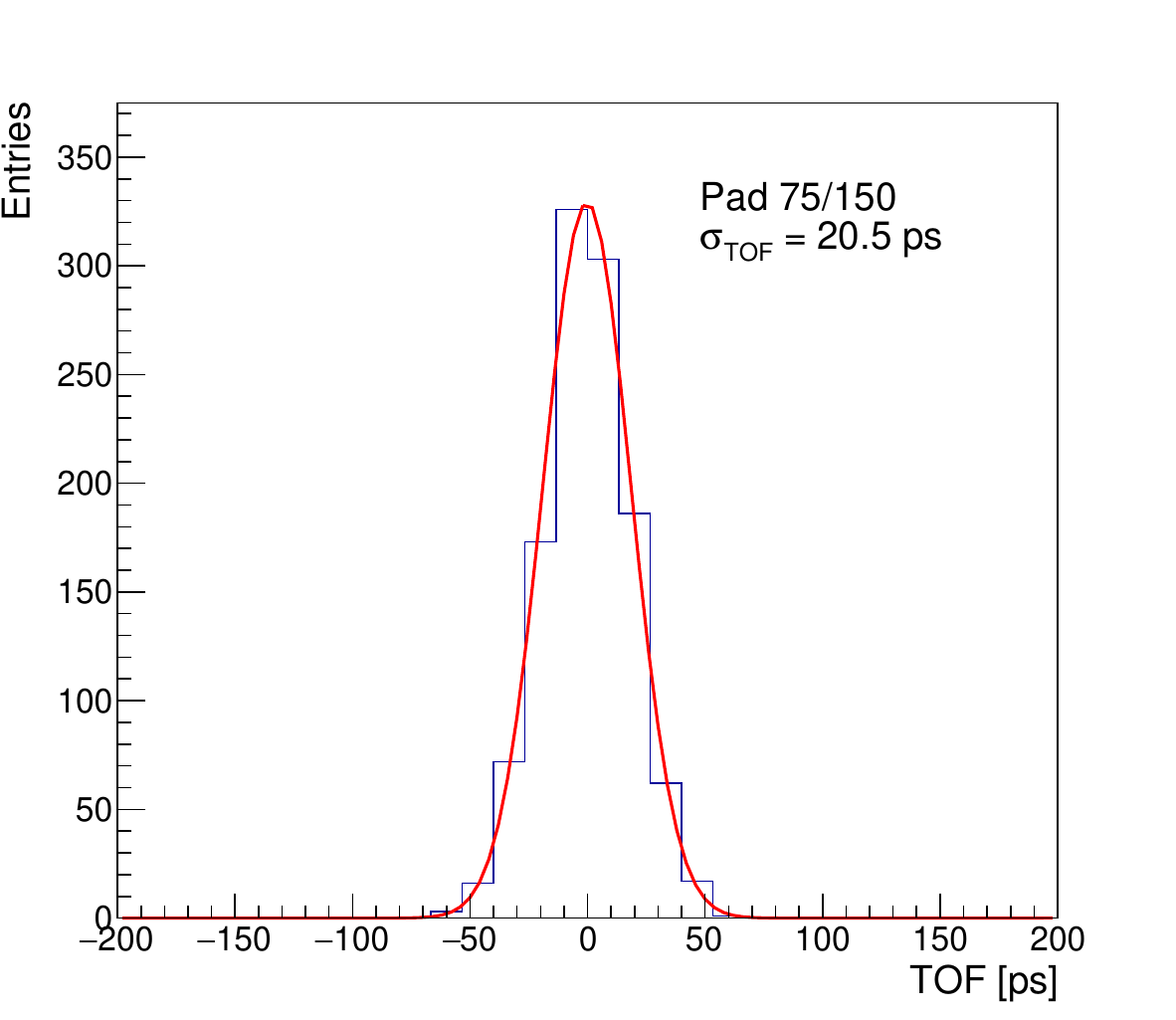} 
        \caption{}
        \label{fig:4channel}
    \end{subfigure}
    \caption{(a) schematic of the setup for time resolution measurements using $\beta$ source. (b) the distribution of time of flight, overlaid with a Gaussian fit.}
    \label{fig:laser_time}
\end{figure}

The AC-LGAD is expected to maintain excellent timing resolution inherited from DC-LGAD. The time resolution under laser incidence for 75/150 pad was analyzed. A $2\times2$ array was connected to USTC multi-channel amplifier board, and the selection of events was consistent with the annotations in figure \ref{fig:2D_pad}. The laser signal is used as trigger, and for the case where only one pad signal is used for timing, all four pads obtain consistent results with a time resolution of approximately 38.6 ps, as shown in Figure \ref{fig:1channel}. By utilizing signals from multiple electrodes and using the amplitude square weighted averaging method in eq (\ref{eq:eqtime}), 
\begin{equation}
t =  \frac{ \sum\limits_i a_i^2 t_i}{ \sum\limits_i  a_i^2 },
\label{eq:eqtime}
\end{equation}
where $a_i$ and $t_i$ are the amplitude and TOA of each pad respectively. The time resolution can be improved to 20.5 ps \cite{c13}.

\begin{figure}[htbp]
    \centering
    \begin{subfigure}[c]{0.46\textwidth}
        \centering
        \includegraphics[width=\textwidth]{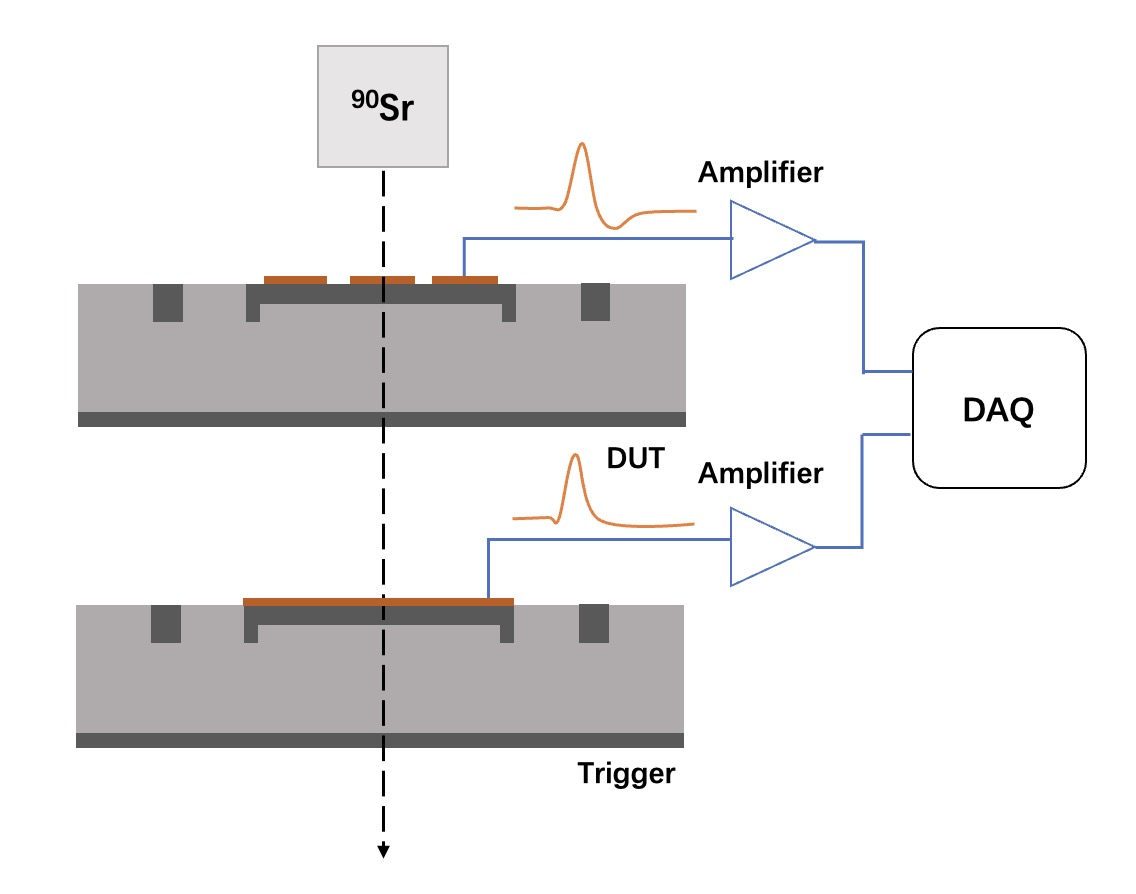} 
        \caption{}
        \label{fig:sub-a}
    \end{subfigure}
    \hfill
    \begin{subfigure}[c]{0.49\textwidth}
        \centering
        \includegraphics[width=\textwidth]{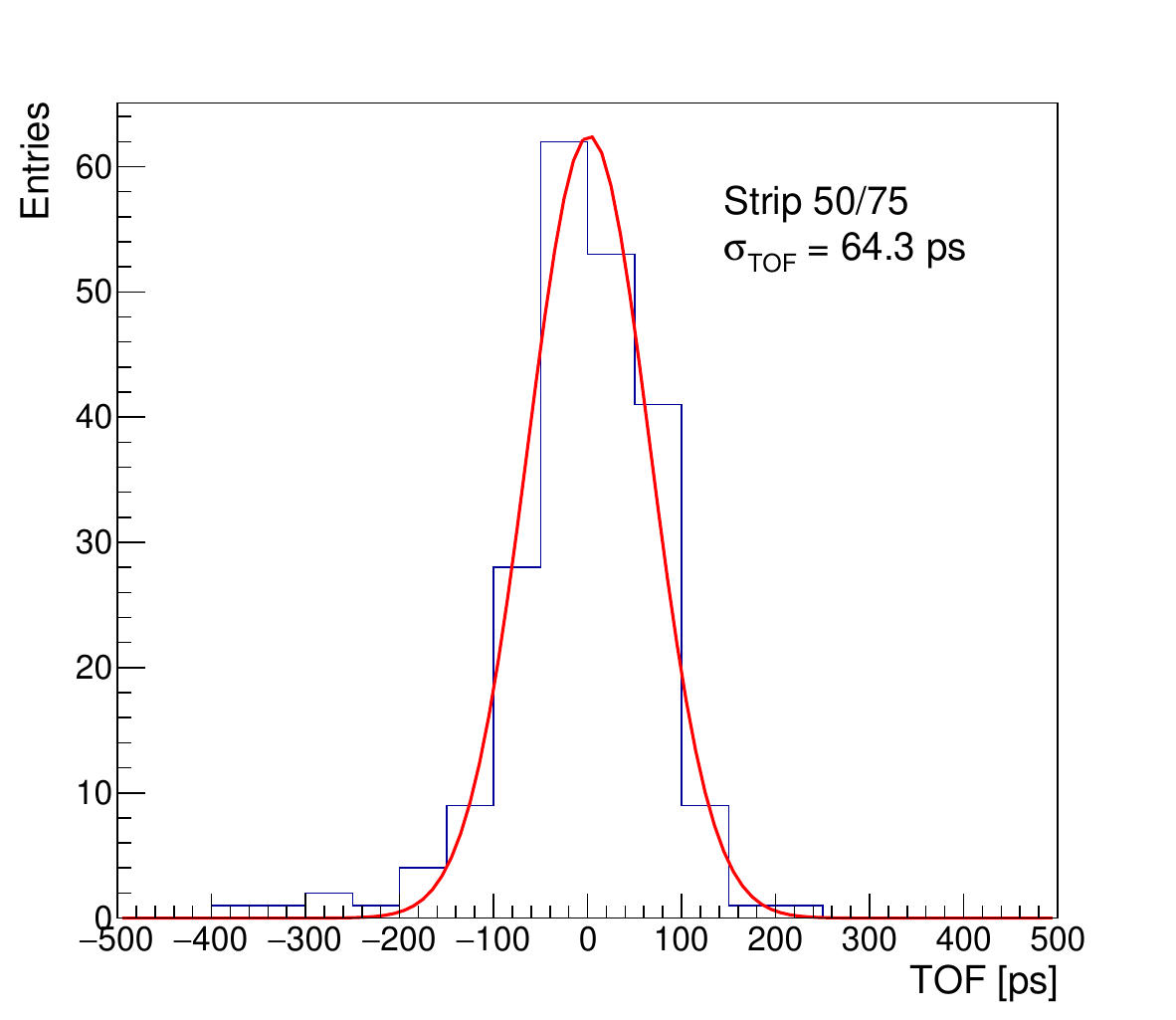} 
        \caption{}
        \label{fig:sub-e}
    \end{subfigure}
    \caption{(a) schematic of the setup for time resolution measurements using $\beta$ source. (b) the distribution of time of flight, overlaid with a Gaussian fit.}
    \label{fig:combined}
\end{figure}

To extract the time resolution performances of USTC AC-LGAD sensor under MIP incidence, a $^{90}$Sr beta source is used to measure the time resolution. The schematic of the measurement system is shown in figure \ref{fig:sub-a}.

A calibrated DC-LGAD with a time resolution of 42.9 ps placed beneath the device under test (DUT) as a reference and trigger detector, while a $^{90}$Sr beta source was placed above the DUT. 3 strips with 50/75 $\mu$m pitch were wire-bonded to USTC multi-channel board to read the induced signal when particles hit the vicinity of the electrode. Signals from the reference and DUT are amplified by the USTC amplifier boards and the outputs are recorded by an oscilloscope. The DUT and the reference detectors are maintained within a temperature chamber regulated at 20$^\circ$C to ensure stable operational conditions.

The Time-of-Arrival (TOA) of the sensor is defined by the constant fraction discrimination (CFD) method at 50\%. The fraction of CFD was varied from 10\% to 90\%, and the corresponding time resolution was found to be close. The timestamps given by the 3 strip electrodes are weighted and averaged using eq (\ref{eq:eqtime}). The time-of-flight (TOF) is defined as the difference in TOA of the reference and DUT. Due to the better temporal resolution performance demonstrated by W5 in laser test, the sensor on W5 underwent beta source test.  The TOF distribution at 150 V in figure \ref{fig:sub-e} is fitted with a Gaussian function with a Gaussian width of 64.3 ps. After removing the contribution of the reference detector using eq (\ref{eq:eq1}), the time resolution of DUT is 47.9 ps.

\begin{equation}
\sigma_\textrm{DUT} = \sqrt{\sigma_\textrm{TOF}^2-\sigma_\textrm{ref}^2} 
\label{eq:eq1}
\end{equation}

The jitter, which is proportional to the noise $N$ and inversely proportional to the slope of the signal, can be calculated using eq (\ref{eq:eq2}). The noise can be determined by the root mean square (rms) of the waveform in a time window faraway without signal pulse, and the slope is defined by the slew rate of the rising edge from 10\%  to 90\%. The influence caused by jitter $\sigma_\textrm{jitter}$ is found to be 24.4 ps.

\begin{equation}
\sigma_\textrm{jitter} = \frac{\textrm{Noise}}{\textrm{Slope}}  
\quad\quad
\textrm{Slope}  = \frac{A_{90\%}-A_{10\%}}{T_{90\%}-T_{10\%}}
\label{eq:eq2}
\end{equation}

The optimal time resolution arising from Landau ionization fluctuations for DC-LGAD sensors of 50 $\mu$m active thickness is expected to be around 30 ps \cite{c12}. Our observed resolution is consistent with the expectations for this configuration.

\section{Conclusion}

In conclusion, the first batch USTC AC-LGAD has been fabricated and characterized. A full characterization of sensors is presented, starting with the simulation of static electronic fields and dynamic signals. Then the electrical tests and dynamic properties of our detectors is described. Measurements with the IR laser TCT revealed that the spatial resolution can be significantly higher than the nominal pitch divided by $\sqrt{12}$. A 150 $\mu$m-pitch pad matrix with 100 $\mu$m pad size has shown a position resolution of 6.20 $\mu$m. For strip electrodes with smaller pitch, the position resolution reaches 3.86 $\mu$m. The time resolution measured with beta source at 20$^\circ$C is about 48 ps.

This work has demonstrated the working principle of improving position resolution of LGAD with resistive readout and the feasibility of fabricating AC-LGAD in-house at USTC. A reliable simulation chain has been established. This will serve as the base of our future development, which is to tailor the AC-LGAD design for specific applications with requirements of various spatial and temporal resolutions. 

\appendix

\acknowledgments
This work was partially carried out at the USTC Center for Micro and Nanoscale Research and Fabrication and partially supported by the National Natural Science Foundation of China (grant No. 11961141014).


\bibliographystyle{JHEP}
\bibliography{biblio.bib}

\end{document}